# Optically detected and radio wave-controlled spin chemistry in flavoproteins

Kun Meng[1,2], Linyan Nie[1,2†], Johannes Berger[3†], Nick R. von Grafenstein[1,2], Christopher Einholz[3], Stefan Weber[3], Lars-Oliver Essen[4], Roberto Rizzato[1,2], Erik Schleicher[3], Dominik B. Bucher[1,2*]

[1] Technical University of Munich, TUM School of Natural Sciences, Chemistry Department, Lichtenbergstr. 4,85748 Garching bei München, Germany
[2] Munich Center for Quantum Science and Technology (MCQST), Schellingstr. 4, 80799, München, Germany
[3] Albert Ludwigs-Universität Freiburg, Institute of Physical Chemistry, 79104 Freiburg, Germany
[4] Marburg University, Chemistry Department, 35032 Marburg, Germany

†These authors contributed equally to this work
*Corresponding author: dominik.bucher@tum.de

## Abstract

Optically addressable spin systems, such as nitrogen-vacancy centers in diamond, have been widely studied for quantum sensing applications. In this work, we demonstrate that certain flavoproteins — specifically cryptochrome and iLOV — which generate spin correlated radical pairs upon optical excitation, also exhibit optically detected magnetic resonance (ODMR). Remarkably, the iLOV protein, commonly used in cellular imaging, displays ODMR contrast approaching 50%. We present initial applications including widefield magnetic field sensing and spatial modulation of photoluminescence using radiofrequency pulses and magnetic field gradients. Our results establish radical pairs in proteins as a novel platform for optically addressable spin systems, offering the key advantages of molecular designability and genetic encodability. Moreover, due to the spin-selective nature of radical pair chemistry, the results lay the groundwork for future radiofrequency-based manipulation of biological systems.

## Introduction

Solid-state spins in semiconductors, such as the nitrogen-vacancy (NV) center in diamond[1] or boron vacancy in hexagonal boron nitride[2], are the workhorses of quantum sensing[3] (Fig. 1a). Their spin-photon interface enables optically detected magnetic resonance (ODMR) for various sensing applications in biology and physics. Despite their great properties, synthetic tunability of their optical and spin properties or deterministic fabrication methods have remained elusive[4]. In contrast, molecular spin systems offer a compelling alternative, allowing bottom-up design with two key advantages: i) tunability through precise atomistic control over the molecular structure and its associated properties, and ii) scalability through chemical assembly. Recently, the synthesis of organometallic molecules with an optical spin interface has been demonstrated[5]. In our work, we show that certain flavoproteins also exhibit optically addressable spin states. The main advantage of the genetically encoded protein scaffolds is their biocompatibility and their tunability via rational design[6] or directed evolution[7].
Photoactive flavin-containing proteins, such as cryptochromes (Crys) and light–oxygen–voltage (LOV) domain proteins, respond to blue light by generating spin correlated radical pairs (SCRPs) involving the flavin cofactor and a nearby amino acid residue (Fig. 1b). These radical pairs are initially formed in a distinct spin-correlated singlet or triplet state[8]. Spin interconversion between these states, governed by anisotropic hyperfine interactions and influenced by external magnetic fields, can modulate the recombination kinetics and ultimately affect the photochemical outcomes of the reaction[9]. In LOV proteins, these SCRPs only form if the natural reaction is blocked by a mutation, in which case they are not involved in signaling[10,11]. Crys, however, are believed to play a direct role in magnetoreception, which enables birds, for example, to navigate using the Earth's magnetic field[8,12,13].
In parallel to our work, ODMR in engineered LOV proteins (MAGLOV)[14], in red fluorescent protein-flavin systems[15] and in fluorescent protein systems[16] — the latter of which does not rely on SCRPs have been shown. Our work shows that radical pair–based ODMR also occurs in the archetypal cryptochromes[17] as well as in iLOV[18], a well established fluorescent reporter.

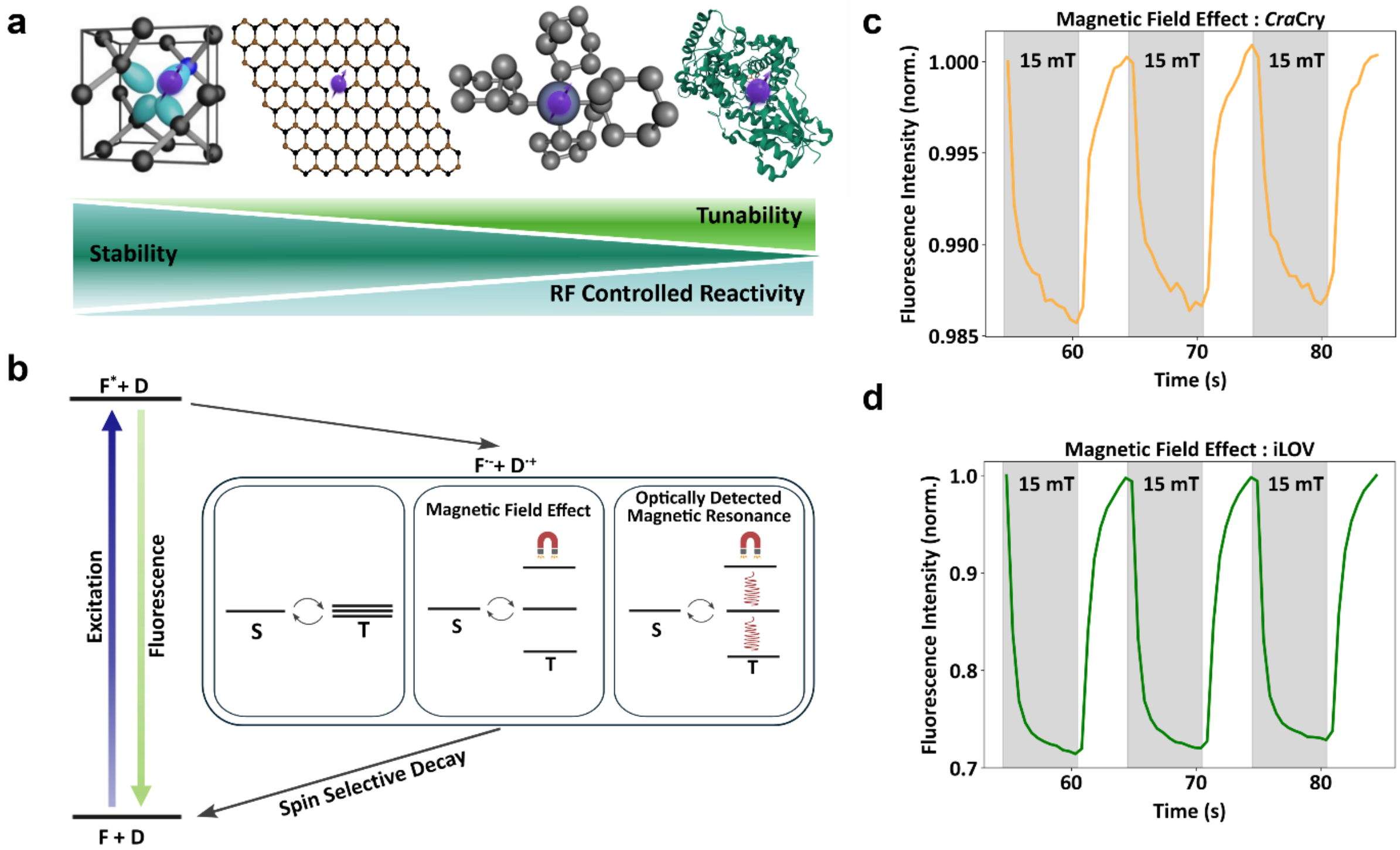

**Figure 1. Optically addressable spin systems. a)** Overview of various spin systems, including the nitrogen-vacancy center in diamond, the boron vacancy in hexagonal boron nitride, organo-metallic molecules, and, as demonstrated in this work, proteins. The degree of tunability increases from semiconductor-based systems to biological systems. Radical pairs and their yields in chemical and biological systems can be controlled by radio wave (RF) fields. **b)** Simplified schematic of the photophysical and spin dynamics of a photoinduced spin-correlated radical pair (SCRP) based on flavin (F) and a donor (D). The interconversion between triplet and singlet states can be influenced by magnetic fields (magnetic field effect, MFE) and/or RF frequencies (optically detected magnetic resonance, ODMR), which decay through spin state selective channels. **c)** Magnetic field effect (~1.5%) observed in the cryptochrome from the green alga *Chlamydomonas reinhardti (Cra*Cry*)*. **d)** Magnetic field effect (~30%) observed in improved Light-oxygen-voltage-sensing protein (iLOV).

## Results and Discussion

**Magnetic field effects.** In the first step, we measure the photoluminescence (PL) of the animal-like Cry from the green alga *Chlamydomonas reinhardti (Cra*Cry*)* as a function of the applied magnetic field strength (Earth's magnetic field and 15 mT) (Fig. 1c). We observe that the PL intensity is affected by the presence of a magnetic field — a phenomenon known as magnetic field effect (MFE). This effect can be attributed to the modulation of the singlet-triplet interconversion of the formed SCRP[12,19–21] (Fig. 1b), which alters the concentration of the different flavin species (mainly the oxidized flavin) under the given experimental equilibrium conditions. Details and reference experiments can be found in Supplementary Note 1 and 2. We also include a LOV protein – specifically iLOV – in our studies, which is a domain that has been optimized to function as a fluorescent marker in cellular imaging[18,22]. Similarly, we observe a MFE, albeit with greater contrast (Fig. 1d) which is laser excitation power dependent (Supplementary Note 3).

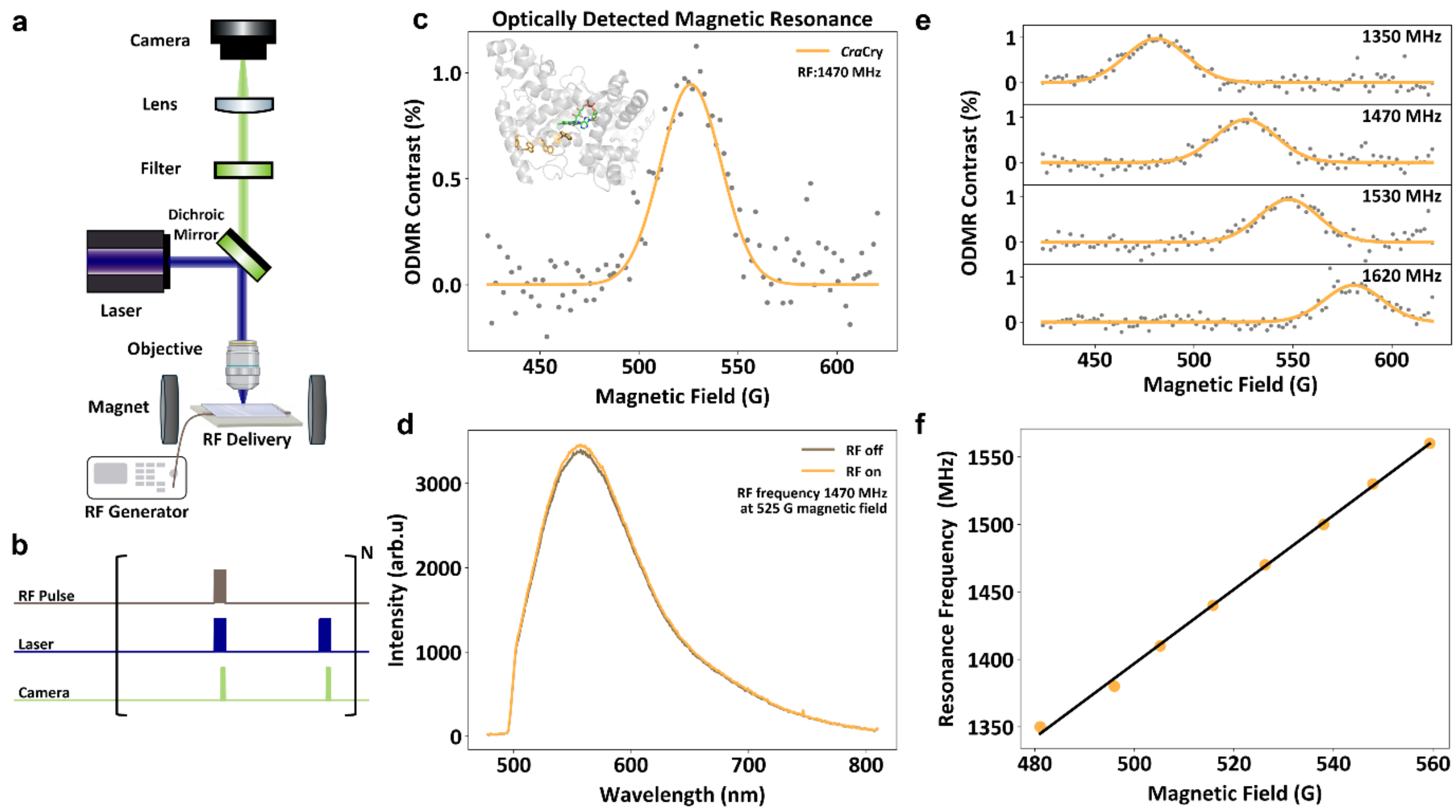

**Figure 2. Optically detected magnetic resonance of *Cra*Cry. a)** Experimental setup of the ODMR experiments. **b)** Continuous ODMR pulse sequence. The pauses between successive experiments provide time for sample recovery. **c)** ODMR spectrum of *Cra*Cry recorded at an excitation frequency of 1470 MHz. *Cra*Cry crystal structure including the FAD cofactor and the electron-transfer chain (orange) is shown in the inset (pdb 5ZM0). **d)** Difference of the *Cra*Cry photoluminescence spectra under RF irradiation. **e)** ODMR spectra of *Cra*Cry recorded under different RF excitation frequencies. **f)** The dependence of the *Cra*Cry resonance frequency on magnetic field strength is consistent with a $g \approx 2$ spin system.

**Optically detected magnetic resonance of *Cra*Cry.** While MFEs have been studied in detail previously[12,20,23], it inspired us to perform ODMR experiments, similar to their optically active solid-state spin counterparts[24]. In these experiments, radio or microwave (RF) frequencies are used to address spin transitions that couple to the optical transitions and can be read out from the PL intensity. Therefore, we developed a modified ODMR setup[25] that includes temperature-controlled sample handling, a tunable magnetic field $B_0$, and electronics for RF delivery, combined with sensitive PL detection (see Materials and Methods) (Fig. 2a). In these experiments, PL intensity is detected as a function of magnetic field. A typical ODMR pulse sequence is shown in Figure 2b, where we apply a specific RF frequency and simultaneously record the PL intensity. To cancel noise (e.g., laser intensity fluctuations), a reference measurement is recorded without the RF pulse. Typically, in ODMR experiments the magnetic field is held constant while the RF frequency is swept[25]. We avoid this approach here, as it can lead to unwanted modulation of the resonance line shape due to the frequency-dependent characteristics of the RF delivery structures (Supplementary Note 4). Instead, we keep the RF frequency fixed and sweep the magnetic field strength. We use a previously characterized solid-state spin system ($S$=1/2) in boron nitride nanotubes (BNNTs)[26,27] for calibrating the magnetic field strength in our experiment (see Supplementary Note 5).
First, we apply a RF frequency of 1470 MHz, sweep the magnetic field from 425 G to 620 G and monitor the PL intensity of the *Cra*Cry sample. An enhanced PL intensity is observed around 525 G, corresponding to the expected spin transition of a free electron (Fig. 2c and d). To verify that the signal indeed originates from an electron spin resonance, we excite the *Cra*Cry sample at different RF frequencies and sweep the magnetic field strength in each case (Fig. 2e). The ODMR resonance shifts consistently according to a $g \approx 2$ system, confirming that we are observing an electronic spin resonance (Fig. 2f).

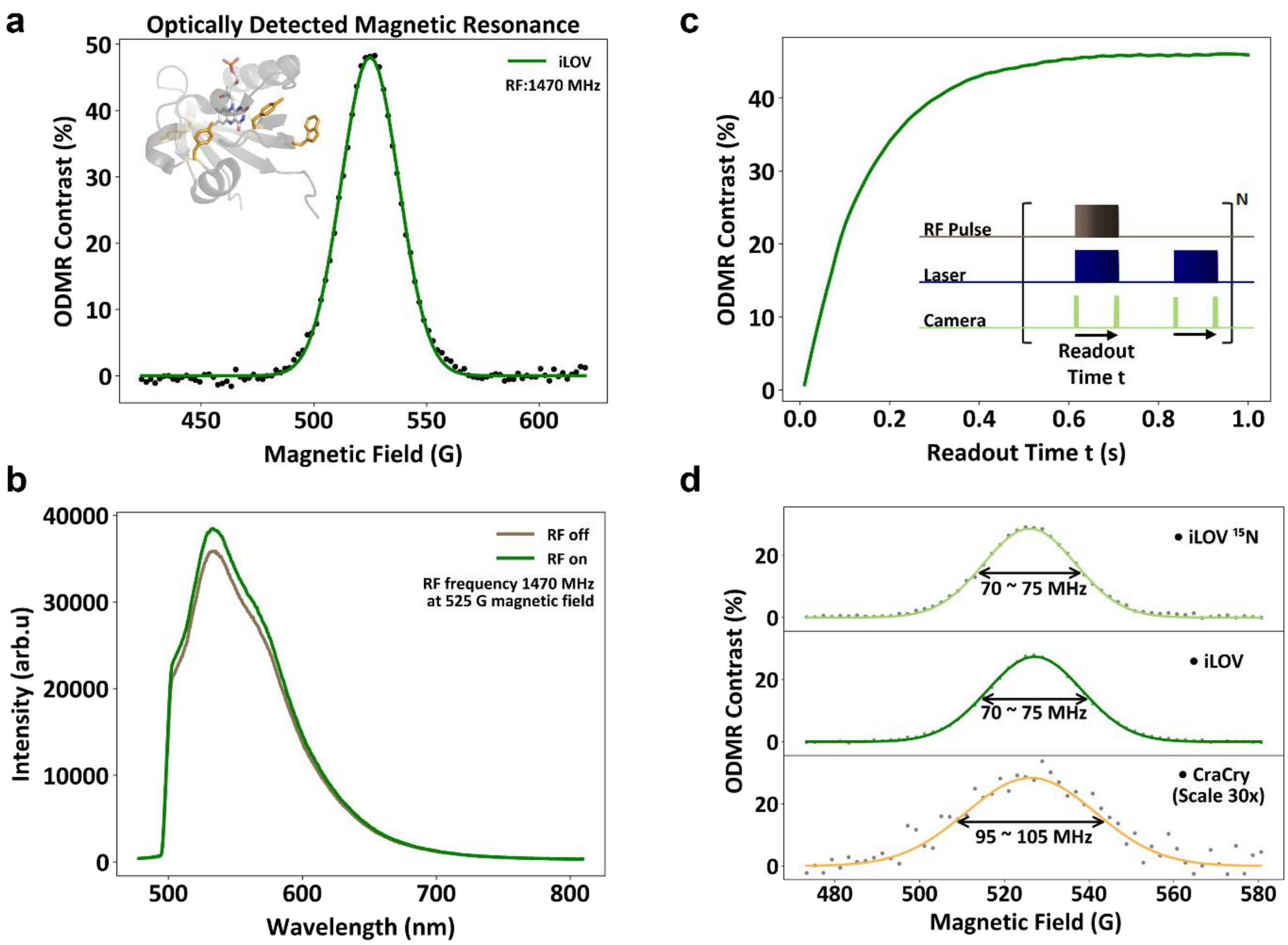

**Figure 3. Optically detected magnetic resonance of iLOV. a)** ODMR spectrum of iLOV recorded at a RF excitation frequency of 1470 MHz. iLOV crystal (pdb 4EET) structure including the FMN cofactor and aromatic amino acids (orange) is shown in the inset. **b)** Difference of the iLOV photoluminescence spectra under RF irradiation. **c)** The ODMR contrast as a function of the readout time indicates a slow buildup, reflecting slow underlying chemical kinetics. **d)** Comparison of linewidths of uniformly labeled $^{15}$N iLOV, iLOV and *Cra*Cry.

**Optically detected magnetic resonance of iLOV.** In the following, we perform ODMR experiments on iLOV. The observed ODMR signals are qualitatively similar to those of *Cra*Cry, but with an impressive contrast of nearly 50% after optimization (Fig. 3a and Supplementary Note 6). In this case as well, we confirm a $g \approx 2$ system (Supplementary Note 7). We also note the drastically enhanced brightness and stability of iLOV compared to *Cra*Cry (Supplementary Note 8). PL spectra indicate an increase in the $FMN_{ox}$ state upon RF excitation (Fig. 3b). Importantly, the signal builds up over time (Fig. 3c), indicating that — similar to the MFE (Fig. 1c and 1d) — the ODMR contrast arises from a slow shift in the chemical equilibrium of the flavin states driven by spin chemistry[9,12]. The dynamics are governed by excitation and RF power as well as other factors influencing the chemical equilibrium. The observation is best explained by the formation of long-lived states, for example generated through redox or (de)protonation reactions, which subsequently decay on the timescale of milliseconds to seconds. Although these states may be only weakly populated after optical excitation, their accumulation under continuous optical and RF excitation conditions leads to a substantial ODMR contrast. In addition, we perform pulsed ODMR experiments in which optical excitation and RF manipulation are temporally separated (Supplementary Note 9). These measurements demonstrate that RF manipulation is possible during the lifetime of the flavin triplet state observed in transient absorption experiments, which is also the timescale on which SCRPs are being formed. All of these observations point toward the SCRP mechanism[8] as a tentative explanation for the observed ODMR effect (Fig. 1b). In the MFE experiments, the applied magnetic field splits the triplet states, which influences the singlet–triplet interconversion. This alters the recombination kinetics of the radical pair, ultimately affecting the equilibrium concentration of flavin molecules in the ground state under continuous optical excitation. The applied RF fields induce spin transitions between the $T^0$ and $T^+/T^-$ triplet states and thus reverse the MFE (Fig. 1c and d), leading to an increased PL intensity.
We note that ODMR can also originate directly from triplet states[16,28]. However, such a mechanism typically does not generate long-lived states on the millisecond to second timescale and, more importantly, would be expected to produce a pronounced zero-field splitting in the GHz range[29], which is not observed in our data.

**Linewidth comparison.** Next, we examine the ODMR lineshapes, which in SCRPs may be influenced by hyperfine couplings as well as exchange and dipolar interactions, among other factors. To investigate the role of hyperfine couplings, we produce a uniformly labeled $^{15}N$ iLOV variant (see Materials and Methods for details). Interestingly, we observe very similar linewidths (~70 MHz) in both the natural-abundance and isotopically labeled variants, indicating that the linewidth of iLOV is not predominantly limited by interactions with $^{15}N$ nuclear spins. We measure the linewidth as a function of RF power to avoid power broadening (see details in Supplementary Note 10). Moreover, the *Cra*Cry sample exhibits a significantly broader resonance linewidth (~100 MHz) compared to iLOV (~70 MHz) under identical RF drive conditions. This result is surprising, as the radical pair separation in iLOV (~ 12 kDa) is anticipated to be smaller, leading to stronger exchange and dipolar interactions compared to the *Cra*Cry sample (~ 58 kDA). At present, we do not have a definitive explanation for this behavior. However, our data may indicate either that the observed ODMR of *Cra*Cry does not originate from the previously identified SCRP[12], or that a distribution of distinct SCRP configurations leads to a broadening of the line.

**Magnetic field sensing.** The ability to detect ODMR in proteins opens up a wide range of potential applications. Similar to optically active defects in solid-state systems, spin resonance can be used for magnetic field sensing[24]. Due to its superior ODMR performance, brightness, robustness, compact size, and suitability for biotechnological applications, iLOV was selected for subsequent experiments. To demonstrate this, we position a small magnet adjacent to the sample, generating a magnetic field gradient across the microscope's field of view (Fig. 4a). Unlike in our previous ODMR experiments, we retain spatial information by analyzing the spectrum recorded at each camera pixel. The resulting ODMR resonance shifts across the field of view reflect the local magnetic field variations. In these experiments, we record the ODMR as a function of RF frequency. Although the detailed lineshape and structure of the ODMR signal is affected by the RF delivery, we empirically found that weighting the ODMR contrast allows extraction of an average resonance frequency, which serves as a reliable calibration metric for the magnetic field (Supplementary Note 12). Applying this fit pixel-wise allows us to construct a magnetic field map, visualizing the gradient across the sample (Fig. 4b). We validate the magnetic field gradient with the BNNTs solid-state spin system (Supplementary Note 13). This technique offers a promising route for genetically encoded *in vivo* cell magnetic field sensing, providing a complementary approach to quantum diamond microscopy, which is often limited by the distance between the sensor and the biological sample[30].

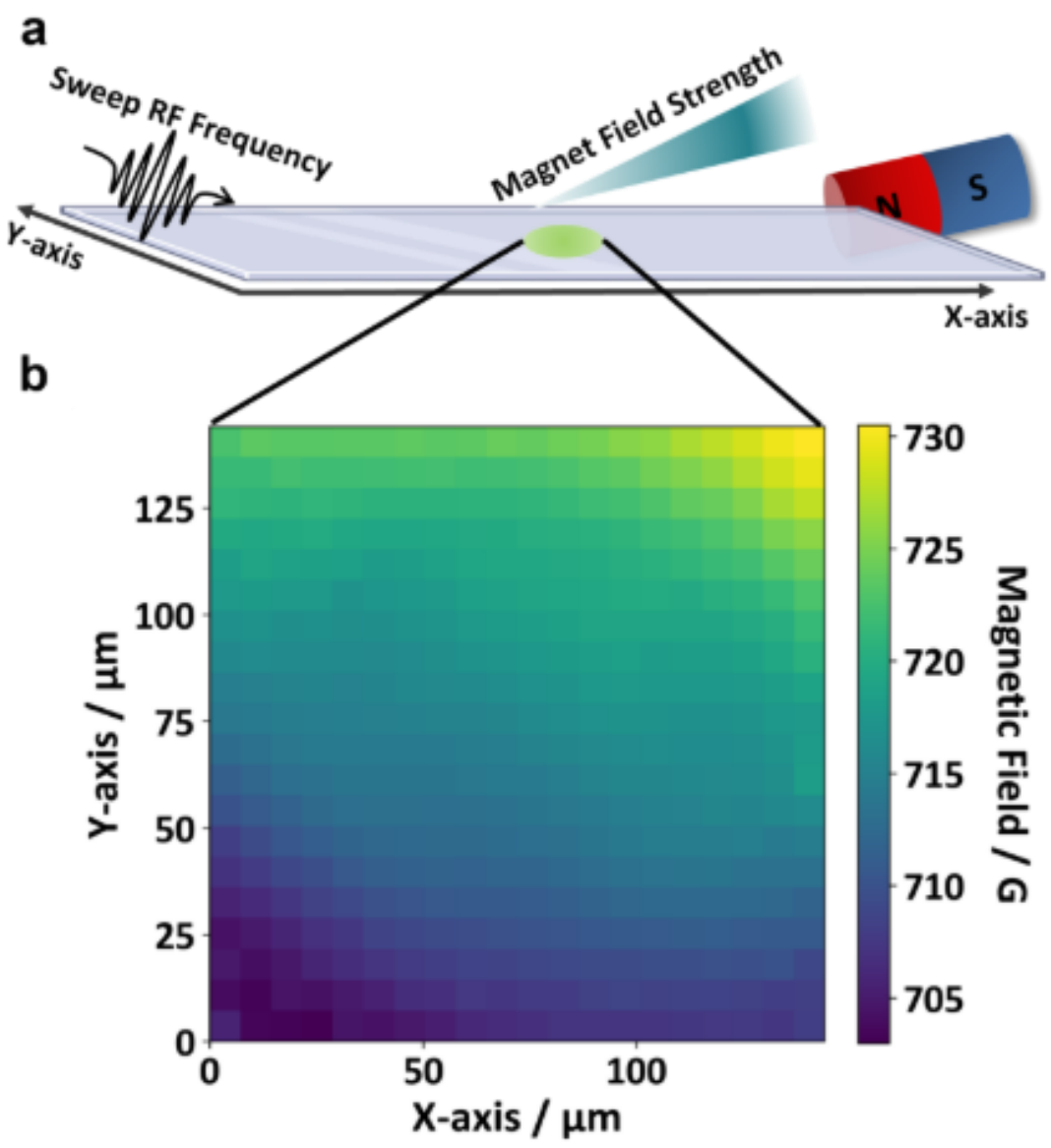

**Figure 4. Magnetic field imaging using iLOV. a)** Schematic of the experimental setup for magnetic field sensing. A small magnet placed next to the iLOV sample generates a magnetic field gradient across the microscope's field of view. **b)** Fitting the ODMR spectra pixel-wise, a spatial map of magnetic field strength is obtained, visualizing the gradient across the sample.

**Spatial RF control.** Finally, the combination of RF control and magnetic field gradients enables spatial control of the radical pair in the proteins. Analogous to techniques used in magnetic resonance imaging, a magnetic field gradient *B(x)* allows to encode the position into a resonance frequency according to the relation $f = \gamma B(x)$ , where $\gamma$ is the gyromagnetic ratio of the electron (Fig. 5a and b). By applying a specific RF frequency $f_{RF}$, only the region — or "slice" — that satisfies this resonance condition is selectively excited. This localized excitation by RF pulses alters the radical recombination yield and, consequently, the PL at that position. Spatial control of PL is visualized by calculating the PL ratio between measurements with RF applied (RF on) and without RF (RF off), which is shown for different frequencies $f_{RF}$ in Figure 5d. By tuning the RF frequency, we can selectively address proteins in certain regions, where the spatial resolution is determined by the strength of the applied magnetic field gradient.

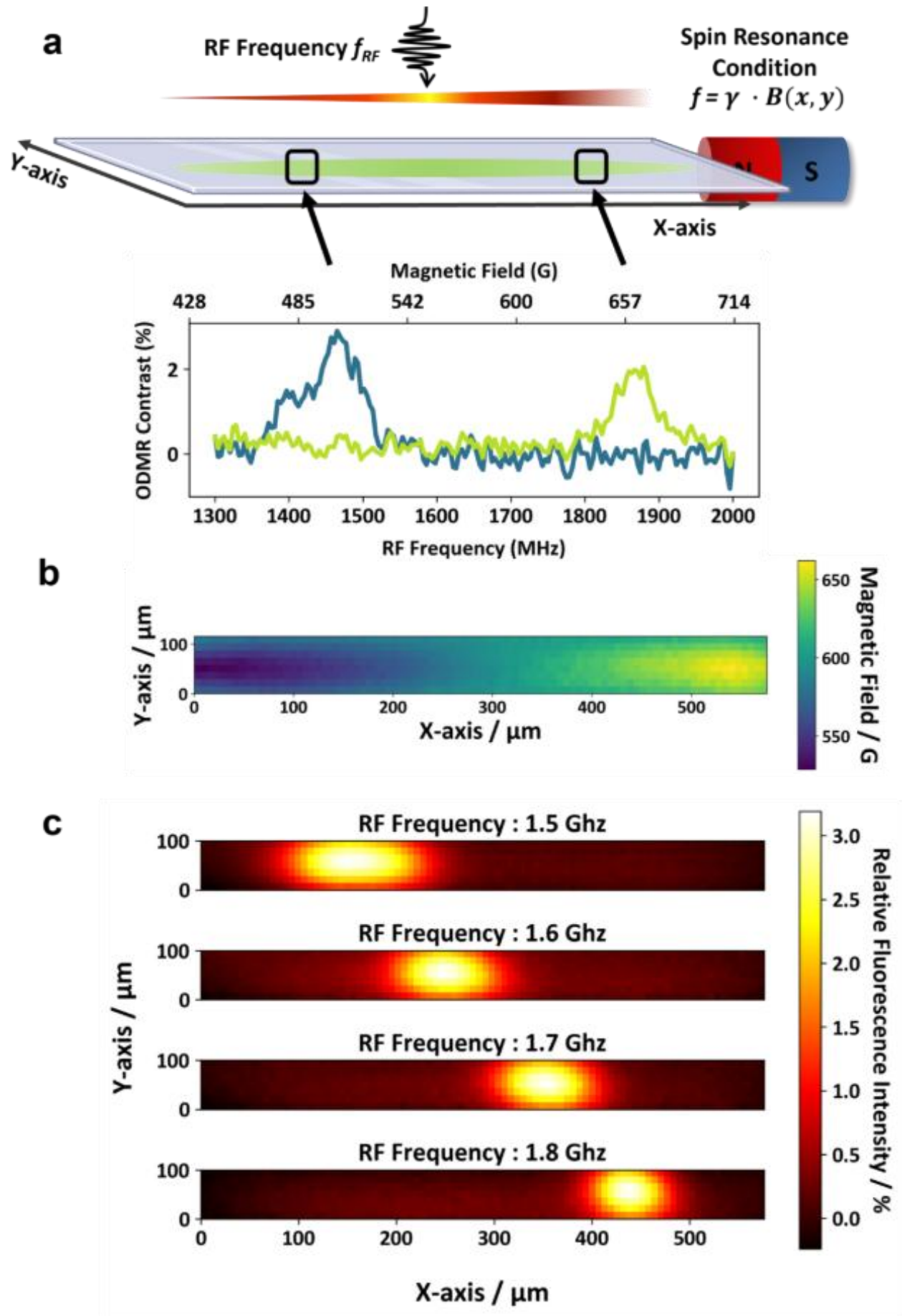

**Figure 5. Spatial photoluminescence (PL) control based on the iLOV spin resonance. a)** The magnetic field gradient, combined with RF frequencies, enables spatially selective excitation of regions that match the spin resonance condition. Depending on the position, the protein has a different ODMR resonance, encoded in the magnetic field gradient shown in **b)**. **c)** PL detection as a function of RF frequency demonstrates spatial control over PL intensity, and consequently, over spin-dependent radical recombination processes. Each applied RF frequency *f* selectively excites iLOV spin transitions (see a)) at specific locations that match the ODMR resonance condition, resulting in an increase in PL. The image shows the PL ratio with RF on divided by RF off.

On the one hand, the ability to modulate the PL intensity of proteins can enhance sensitivity — for example, by suppressing background PL[31] — or enable super-resolution techniques in which spatial resolution is encoded in magnetic field gradients, thereby surpassing the limits of optical resolution[32]. On the other hand, our results demonstrate that (photo)chemistry of flavin states can be controlled using RF fields — a key distinction from solid-state spin defects, where the ODMR response is governed by photophysical processes (Fig. 1a). This paves the way toward future RF-based control of biological processes such as gene expression or signaling[33–37].

In summary, we have shown that spin states in *Cra*Cry and iLOV proteins can be manipulated with RF fields and read out optically at ambient temperature. The inherent tunability of proteins through rational design[6] or directed evolution[7] offers exciting opportunities for engineering spin-optical interfaces. We anticipate broad applications across quantum technologies, bioimaging and RF controlled biochemistry[33–36].

**Methods**

**Sample preparation.** All proteins were produced according to Paulus et al.[38] Uniformly labeled $^{15}$N isotope iLOV protein were produced according to Mengli Cai et al.[39], and purified according to Günther et al[40]. *Cra*Cry samples were prepared in a buffer containing 50 mM $NaH_2PO_4$ pH 7.8, 100 mM NaCl, and 50% (v/v) glycerol. iLOV samples were prepared in a buffer containing 50 mM HEPES, 100 mM NaCl, 20% (v/v) glycerol at pH 7.0. Prior to experiments, 20 μL of the protein sample was transferred to a quartz cuvette (Z805963, Hellma Analytics) and sealed with a 12 mm round cover glass (W07-4615002, Labstellar). Due to the difference in PL brightness, the experiments were performed using *Cra*Cry and iLOV at concentrations of ~200 μM and ~20 μM, respectively.

**Optical setup.** A 447 nm blue laser (iBEAM-SMART-445-S_14935, TOPTICA Photonics AG) was directed through a dichroic mirror (MD498, Thorlabs) and focused onto the sample using a 50x objective (MXPLFLN, Olympus). A 180 mm tube lens (TTL180-A, Thorlabs) was paired with the objective to achieve optimal magnification. Upon excitation, green PL emitted from the samples was collected by the same objective, passed through the dichroic mirror and a 525 nm bandpass filter (MF525-39, Thorlabs), and imaged using a sCMOS camera (Kinetix, Teledyne Photometrics).

**Magnetic field effect experiments.** The quartz cuvette was positioned on an aluminum cage plate. To prevent protein thermal degradation, temperature control was implemented using a single-stage Peltier element (TECF2S, Thorlabs) positioned beneath the aluminum cage plate, maintaining the sample at 15 °C. A custom-built copper coil provided a modulated magnetic field (0 - 15 mT) at the sample position. Under continuous 447 nm laser illumination, PL intensity was recorded using the camera with 500 ms exposure time. A custom Python script was used to synchronize the magnetic field modulation with image acquisition.

**RF strip line antenna design.** RF strip line structure was designed in-house and fabricated (CERcuits, Geel) to provide RF fields for coherent electron spin manipulation. The strip line structure is implemented on a PCB with outer dimensions of 50.6 mm × 40.6 mm. Aluminum oxide was chosen as substrate material with a thickness of 500 μm, due to its high thermal conductivity. The conductive layer and ground plane are both composed of copper, each with a thickness of 70 μm. The signal conductor was implemented as a straight line with a width of 480 μm, resulting in a characteristic impedance of ~ 50 Ω. A ~ 2 mm-long constriction was introduced at the center of the line to reduce the width to 240 μm, thus enhance the local RF field strength in the sample region. This narrowed region is designed to boost current density locally, which in turn increases RF magnetic field in the vicinity of the sample positioned above it. The microstrip layout was optimized to maintain impedance matching and minimize signal reflection, while also providing sufficient field strength for fast, coherent spin-state manipulation.

**Optically detected magnetic resonance experiments.** The experimental setup employed a Pulse Streamer 8/2 (Swabian Instruments) for precise synchronization of the laser illumination, RF, and camera imaging acquisition. A custom lab software (QuPyt, https://github.com/KarDB/QuPyt) was used for precise timing. RF frequencies were generated using a signal generator (R&S®SMB100A, Rohde & Schwarz) set to an output power of -5 dBm. The pulses were generated by a RF switch (ZASWA-2-50DRA+, Mini-Circuits), amplified with a broadband RF amplifier (KU PA BB 070270-80 B, Kuhne electronic), and delivered to the sample via an RF strip line antenna. The antenna was terminated with a 50 Ω high-power load (TERM-50W-183S+, 50 W, SMA-M, Mini-Circuits) to prevent reflections. The sample in the cuvette was placed on the RF strip line. The strip line was placed on the aluminum cage plate and the temperature of sample was kept at 15 °C. This configuration provided both effective heat dissipation and stable RF delivery.

The measurement protocol employed a synchronized cycle consisting of a 500 ms RF pulse with simultaneous laser illumination and 100 ms camera exposure ending synchronously with the RF and laser pulses, followed by a 9.5 s recovery period and a reference measurement with laser illumination and camera exposure without RF.

A permanent magnet placed beneath the sample plate provided a static magnetic bias. In addition, a copper coil was positioned at the sample site. As calibrated in Supplementary Note 5, this dual-magnet configuration enabled sweeping of the magnetic field from 425 G to 620 G. At each RF frequency, we swept the magnetic field across this range and monitored the corresponding change in PL intensity. PL intensity was determined as the mean photo counts value within a defined 100x100 pixel area (~ 13x13 μm). Typical ODMR experiment durations were between 0.5 and 2 hours.

**Magnetic field imaging experiments.** In Figure 4, wide-field illumination was implemented by expanding the laser beam with an additional lens in front of the 10× objective (10X Mitutoyo Plan-Apochromat Objective, Thorlabs), producing approximately 150 μm x 150 μm field-of-view. The magnet used for the ODMR measurements was removed, and a small magnet was positioned adjacent to the sample. For measuring the local magnetic field, we swept the RF frequency to determine the ODMR resonance. To achieve a higher signal-to-noise ratio (SNR), the image was binned by every 10x10 pixels. For each binned pixel, the resonant frequency and corresponding magnetic field were extracted from the ODMR spectrum by applying the procedure in Supplementary Note 12.

**RF control experiments.** In Figure 5, the laser beam was focused using a cylindrical lens (LJ1653RM, Thorlabs) to generate an illumination area of approximately 500 μm along the x-axis and 100 μm along the y-axis. The small magnet was placed to generate a large magnetic field gradient along x-axis. The RF frequency was swept across an 800 MHz range while PL intensity changes were monitored pixel-wise. The image was binned by every 10 pixels along the x-axis and every 20 pixels along the y-axis. For each binned pixel, the resonant frequency and corresponding magnetic field were extracted from the ODMR spectrum by applying the procedure in Supplementary Note 12. Spin-selective excitation experiments were performed under the same magnetic field gradient. The sample was exposed to fixed RF frequencies of 1.5, 1.6, 1.7, and 1.8 GHz. At each frequency, the PL signal was averaged over 50 measurements and normalized to a reference measurement taken without RF excitation.

**Data analysis.** The MFE was corrected by subtracting the photobleaching decay, which was fitted with a linear curve over the selected time range. The ODMR contrast was calculated by first dividing the PL intensity measured with RF on by the PL intensity with RF off. This ratio was then corrected by subtracting the baseline. The data were fitted with a Gaussian function.

## Acknowledgements

**Funding.** This project has been funded by the Bayerisches Staatministerium für Wissenschaft und Kunst through project IQSense via the Munich Quantum Valley (MQV), the Deutsche Forschungsgemeinschaft (DFG, German Research Foundation)—412351169 within the Emmy Noether program and the European Research Council (ERC) under the European Union's Horizon 2020 research and innovation programme (Grant Agreement No. 948049). The authors acknowledge support and seed funding by the DFG under Germany's Excellence Strategy–EXC 2089/1-390776260 and the EXC-2111 390814868. E.S. thanks the Hans Fischer Gesellschaft for continuous support.

**Author Contributions.** D.B.B. conceived the study. K. M. designed and built the setup and performed the experiments. L.N. took care of the sample handling. J.B., S.W., and L.E. provided the samples. N.R.v.G. designed the RF antenna. K.M., R. R., C. E., J. B., L.N., E.S., and D.B.B. analyzed and discussed the data. R. R. helped setting up the experiment and implemented the BNNT experiments. E.S. and D.B.B. supervised the study. D.B.B. wrote the manuscript with input from all authors.

## References

1. Schirhagl, R., Chang, K., Loretz, M. & Degen, C. L. Nitrogen-Vacancy Centers in Diamond: Nanoscale Sensors for Physics and Biology. *Annu. Rev. Phys. Chem.* **65**, 83–105 (2014).
2. Vaidya, S., Gao, X., Dikshit, S., Aharonovich, I. & Li, T. Quantum sensing and imaging with spin defects in hexagonal boron nitride. *Advances in Physics: X* **8**, 2206049 (2023).
3. Degen, C. L., Reinhard, F. & Cappellaro, P. Quantum sensing. *Rev. Mod. Phys.* **89**, 035002 (2017).
4. Mani, T. Molecular qubits based on photogenerated spin-correlated radical pairs for quantum sensing. *Chemical Physics Reviews* **3**, 021301 (2022).
5. Bayliss, S. L. *et al.* Optically addressable molecular spins for quantum information processing. *Science* **370**, 1309–1312 (2020).
6. Khakzad, H. *et al.* A new age in protein design empowered by deep learning. *Cell Systems* **14**, 925–939 (2023).
7. Packer, M. S. & Liu, D. R. Methods for the directed evolution of proteins. *Nat Rev Genet* **16**, 379–394 (2015).
8. Hore, P. J. & Mouritsen, H. The Radical-Pair Mechanism of Magnetoreception. *Annual Review of Biophysics* **45**, 299–344 (2016).
9. Steiner, U. E. & Ulrich, T. Magnetic field effects in chemical kinetics and related phenomena. *Chem. Rev.* **89**, 51–147 (1989).
10. Schleicher, E. *et al.* On the Reaction Mechanism of Adduct Formation in LOV Domains of the Plant Blue-Light Receptor Phototropin. *J. Am. Chem. Soc.* **126**, 11067–11076 (2004).
11. Kay, C. W. M. *et al.* Blue light perception in plants: Detection and characterization of light-induced neutral flavin radical in a C450A mutant of phototropin. *Journal of Biological Chemistry* **278**, 10973–10982 (2003).
12. Kattnig, D. R. *et al.* Chemical amplification of magnetic field effects relevant to avian magnetoreception. *Nature Chem* **8**, 384–391 (2016).
13. Xu, J. *et al.* Magnetic sensitivity of cryptochrome 4 from a migratory songbird. *Nature* **594**, 535–540 (2021).
14. Abrahams, G. *et al.* Quantum spin resonance in engineered proteins for multimodal sensing. *Nature* **649**, 1172–1179 (2026).
15. Burd, S. C. *et al.* Magnetic resonance control of reaction yields through genetically-encoded protein:flavin spin-correlated radicals in a live animal. 2025.02.27.640669 Preprint at https://doi.org/10.1101/2025.02.27.640669 (2025).
16. Feder, J. S. *et al.* A fluorescent-protein spin qubit. *Nature* **645**, 73–79 (2025).
17. Chaves, I. *et al.* The Cryptochromes: Blue Light Photoreceptors in Plants and Animals. *Annual Review of Plant Biology* **62**, 335–364 (2011).
18. Chapman, S. *et al.* The photoreversible fluorescent protein iLOV outperforms GFP as a reporter of plant virus infection. *Proceedings of the National Academy of Sciences* **105**, 20038–20043 (2008).
19. Déjean, V. *et al.* Detection of magnetic field effects by confocal microscopy. *Chem. Sci.* **11**, 7772–7781 (2020).

20. Gravell, J. *et al.* Spectroscopic Characterization of Radical Pair Photochemistry in Nonmigratory Avian Cryptochromes: Magnetic Field Effects in GgCry4a. *J. Am. Chem. Soc.* **147**, 24286–24298 (2025).
21. Maeda, K. *et al.* Magnetically sensitive light-induced reactions in cryptochrome are consistent with its proposed role as a magnetoreceptor. *Proceedings of the National Academy of Sciences* **109**, 4774–4779 (2012).
22. Christie, J. M. *et al.* Structural Tuning of the Fluorescent Protein iLOV for Improved Photostability *. *Journal of Biological Chemistry* **287**, 22295–22304 (2012).
23. Xiang, K. M. *et al.* Mechanism of Giant Magnetic Field Effect in a Red Fluorescent Protein. *J. Am. Chem. Soc.* **147**, 18088–18099 (2025).
24. Levine, E. V. *et al.* Principles and techniques of the quantum diamond microscope. *Nanophotonics* **8**, 1945–1973 (2019).
25. Bucher, D. B. *et al.* Quantum diamond spectrometer for nanoscale NMR and ESR spectroscopy. *Nat Protoc* **14**, 2707–2747 (2019).
26. Gao, X. *et al.* Nanotube spin defects for omnidirectional magnetic field sensing. *Nature Communications* **15**, 7697 (2024).
27. Rizzato, R. *et al.* Quantum sensing with spin defects in boron nitride nanotubes. *Nat Commun* **16**, 11333 (2025).
28. Mena, A. *et al.* Room-Temperature Optically Detected Coherent Control of Molecular Spins. *Phys. Rev. Lett.* **133**, 120801 (2024).
29. Kowalczyk, R. M., Schleicher, E., Bittl, R. & Weber, S. The Photoinduced Triplet of Flavins and Its Protonation States. *J. Am. Chem. Soc.* **126**, 11393–11399 (2004).
30. Le Sage, D. *et al.* Optical magnetic imaging of living cells. *Nature* **496**, 486–489 (2013).
31. Miller, B. S. *et al.* Spin-enhanced nanodiamond biosensing for ultrasensitive diagnostics. *Nature* **587**, 588–593 (2020).
32. Zhang, H., Arai, K., Belthangady, C., Jaskula, J.-C. & Walsworth, R. L. Selective addressing of solid-state spins at the nanoscale via magnetic resonance frequency encoding. *npj Quantum Inf* **3**, 31 (2017).
33. Engels, S. *et al.* Anthropogenic electromagnetic noise disrupts magnetic compass orientation in a migratory bird. *Nature* **509**, 353–356 (2014).
34. Ritz, T., Thalau, P., Phillips, J. B., Wiltschko, R. & Wiltschko, W. Resonance effects indicate a radical-pair mechanism for avian magnetic compass. *Nature* **429**, 177–180 (2004).
35. Zwang, T. J., Tse, E. C. M., Zhong, D. & Barton, J. K. A Compass at Weak Magnetic Fields Using Thymine Dimer Repair. *ACS Cent. Sci.* **4**, 405–412 (2018).
36. Chowdhury, F. T., Denton, M. C. J., Bonser, D. C. & Kattnig, D. R. Quantum Control of Radical-Pair Dynamics beyond Time-Local Optimization. *PRX Quantum* **5**, 020303 (2024).
37. Christie, J. M., Gawthorne, J., Young, G., Fraser, N. J. & Roe, A. J. LOV to BLUF: Flavoprotein Contributions to the Optogenetic Toolkit. *Molecular Plant* **5**, 533–544 (2012).
38. Paulus, B. *et al.* Spectroscopic characterization of radicals and radical pairs in fruit fly cryptochrome – protonated and nonprotonated flavin radical-states. *The FEBS Journal* **282**, 3175–3189 (2015).
39. Cai, M. *et al.* An efficient and cost-effective isotope labeling protocol for proteins expressed in shape Escherichia coli. *J Biomol NMR* **11**, 97–102 (1998).
40. Günter, A. *et al.* Small Protein Domains as Potential Spin Labels for In Vitro, Cellular, and Light-Induced Dipolar EPR Spectroscopy. *J. Am. Chem. Soc.* **147**, 22473–22487 (2025).